# Testing Quality Requirements of a System-of-Systems in the Public Sector - Challenges and Potential Remedies

Jacob Larsson

Capgemini,
Växjö, Sweden

`jacob.larsson@capgemini.com`

Markus Borg, Thomas Olsson

SICS Swedish ICT AB,
Lund, Sweden

`{markus.borg, thomas.olsson}@sics.se`

**Abstract.** Quality requirements is a difficult concept in software projects, and testing software qualities is a well-known challenge. Without proper management of quality requirements, there is an increased risk that the software product under development will not meet the expectations of its future users. In this paper, we share experiences from testing quality requirements when developing a large system-of-systems in the public sector in Sweden. We complement the experience reporting by analyzing documents from the case under study. As a final step, we match the identified challenges with solution proposals from the literature. We report five main challenges covering inadequate requirements engineering and disconnected test managers. Finally, we match the challenges to solutions proposed in the scientific literature, including integrated requirements engineering, the twin peaks model, virtual plumblines, the QUPER model, and architecturally significant requirements. Our experiences are valuable to other large development projects struggling with testing of quality requirements. Furthermore, the report could be used by as input to process improvement activities in the case under study.

**Keywords:** Experience report, quality requirements, software testing, system-of-systems, document analysis.

## 1    Introduction

The importance of software is ever-growing in our modern society, as more and more domains depend on software-intensive systems. In a society increasingly relying on software, the quality aspects of the various software systems become critical [5]. Testing Quality Requirements (QR) is known to be difficult in large software engineering projects. In a large study on challenges related to alignment of Requirements Engineering and Testing (RET), Bjarnason *et al.* report that three out of six studied companies struggle with verifying QRs [8]. Several other researchers also acknowledge the importance of managing QRs [17, 21], and have called for additional research [4, 16]. An understanding of software quality should permeate the entire development process [27]. Quality aspects, sometimes referred to as "-ilities", are typically expressed as QRs (a.k.a. non-functional requirements, in contrast to Functional Requirements (FR)) as part of a project's Requirements Engineering (RE). Example categories of QRs according to ISO/IEC 25010 include reliability, operability, and maintainability [26].

In our previous work on reported general challenges of RET alignment in the public sector [28], *verifying quality requirements* was listed as one of the major challenges. The discussion on QRs was however comparatively short, with the overall conclusion being *"many quality aspects cannot be assessed until the system is in operation"*. We now return to the same project 1.5 years later and report experiences from testing QRs of a system that is now partly in operation. Consequently, the current paper constitutes a follow-up of our previous work, i.e., we return to the same development project at a governmental agency in Sweden: a major integration effort with the goal to establish a System-of-Systems (SoS) for managing EU grants. In this paper, we further strengthen the validity of our conclusions by complementing the experiences reported by archival analysis of requirements documentation. Finally, our report responds to a recent national call for more research on SoS in Sweden [3].

Our main contribution is to report five major challenges in verifying QRs in the case under study, namely: 1) changing RE documentation, 2) test managers' need of domain understanding, QRs are neither 3) quantified nor 4) prioritized frequently enough, and 5) simulation of complex operational states in a test environment. We also propose solutions based on the research literature for the five challenges, including improved commination between requirements engineers, architects and test managers (i.e., extending the Cleland-Huang *et al.*'s "twin peaks" model [11] with the test perspective), and early identification of "test significant QRs" (in line with work on architecturally significant requirements by Chen *et al.* [10]).

The paper is structured as follows: Section 2 introduces previous research on QRs, and particularly related work on testing QRs. Section 3 describes the case company and the relevant development processes applied in the project under study. Section 4 presents the research method and the limitations of our work. Section 5 reports our results and our corresponding interpretations. Finally, Section 6 concludes our paper by summarizing the main takeaways.

## 2    Background and related work

While management of QRs is known to play a crucial role in large software engineering projects, the RE research community has not agreed on a single definition [14]. In this paper, we rely on the definition "QRs describe the non-behavioral aspects of a system, capturing the properties and constraints under which a system must operate".

The remainder of this section presents key research on QRs, as well as reported challenges related to both their specification and verification.

### 2.1 Quality requirements and quality models

Several researchers highlight the importance of QRs during software product development. Simply implementing all functional requirements is not enough to ensure a successful product; Disregarding QRs might lead to a product that is too difficult to use or too expensive to maintain [21]. Moreover, poor management of QRs might lead to project overruns and increased time-to-market [17]. Berntsson Svensson argues that QRs are particularly important to market-driven organizations that release products to an open consumer market [4]. Although the importance of QRs is generally acknowledged, Chung and do Prado report that RE research is dominated by functional requirements [14], whereas Ameller *et al.* [1] report that both FR and QR are considered equally important by architects.

*Software quality models* have been developed to support identification of QRs and to help establish control criteria for quality assurance. Two of the most established models are ISO/IEC 25010 [26] (the successor of ISO/IEC 9126) and FURPS+ (a Hewlett Packard adaptation of the original FURPS model [25]), see Table 1. ISO/IEC 25010 composes software product quality into eight characteristics: functional suitability, performance efficiency, compatibility, usability, reliability, security, maintainability, and portability. Each characteristic is further divided into sub-characteristics. FURPS+ is an acronym representing the quality aspects covered by the model: functionality, usability, reliability, performance, and supportability. The '+' was appended to the original FURPS model to highlight additional quality aspects, and constraints on the product, e.g., physical requirements or implementation requirements.

Previous work highlights QRs as among the most challenging aspects to manage during software projects. Moreover, Chung *et al.* report that development organizations often do not properly acknowledge the significance of QRs [15]. Several researchers try to express the overall challenges of QRs. For example, Glinz claims that the mains issues with QRs originate in their *definition*, *classifications*, and *representation* [24]. Chung *et al.* instead argues that the three biggest challenges with QR originates in their nature of being *subjective*, *relative*, and *interacting* [15].

A number of studies point out more specific challenges of QRs. Ernst and Mylopoulos report that software quality aspects are often first *considered at late stages* of the development lifecycle, typically assessed by a set of measures applied to the final product [23]. Also Cleland-Huang *et al.* notice that QRs typically are *discovered late* in software development projects, and often in an *ad hoc fashion* [13]. Borg *et al.* identified a number of QR challenges in a two-unit case study in Sweden [9]. The main finding was again that QRs are *discovered late, if they were discovered at all*. Furthermore, the authors found that the developers in the two cases under study struggled with *QR specification*; QRs are often specified in vague terms that cannot be verified. Berntsson Svensson [4] summarizes the main challenges of QRs presented in the literature as: 1) QRs are *poorly understood*, 2) QRs are *stated informally*, 3) QRs are often *contradicting*, and 4) QRs are *hard to validate*.

| ISO/IEC 25010 | | FURPS+ | |
| --- | --- | --- | --- |
| Functional suitability | Completeness, Correctness, Appropriateness | **F**unctionality | Capability, Reusability, Security |
| Performance efficiency | Time-behaviour, Resource utilization, Capacity | **U**sability | Aesthetics, consistency, responsiveness etc. |
| Compatibility | Co-existence, Interoperability | **R**eliability | Availability, Failure extent, Predictability etc. |
| Usability | Recognizability, Learnability, Aesthetics etc. | **P**erformance | Speed, Efficiency, Throughput, Scalability etc. |
| Reliability | Maturity, Availability, Fault tolerance, Recoverability | **S**upportability | Testability, Flexibility, Installability etc. |
| Security | Confidentiality, Integrity, Accountability etc. | + | Design, implementation, interface, physical requirements |
| Maintainability | Modularity, Reuseability, Testability etc. | | |
| Portability | Adaptablity, Installability, Replaceability | | |

Table 1. Side by side comparison of ISO/IEC 25010 and FURPS+.

### 2.2 Verifying quality requirements

Even in organizations that acknowledge the value of QRs, testing them is often a problem. Berntsson Svensson *et al.* report in an interview study with 11 companies that practitioners frequently fail to specify QRs in quantifiable formats that can be verified by a testing organization [5]. In an in-depth study of a specific requirements specification from one of the companies, Berntsson Svensson *et al.* find that 56% of the QRs are expressed with a quantified quality level [6]. The authors note however, that quantification is not suitable for all types of QRs, e.g., security. Shahrokni and Feldt address testing of QRs in a safety-critical context (mainly software robustness). They analyze three requirements specifications at a case company, and show that the fraction of quantified QRs varies between 5% and 45% [34]. Also, they report that the number of ambiguous QRs are overrepresented in the set of non-quantified QRs.

In a multi-unit case study by Bjarnason *et al.*, testing QRs is highlighted as one of 16 challenges in RET alignment [8]. Some of the reported root causes confirm findings from previous research, including: 1) specification of testable QRs, and 2) subjectively judging whether an QR has passed or not. Our previous paper confirmed that testing QRs is a major RET challenge also in the current case [28], primarily since most quality aspects cannot be assessed until the full system under development is in operation.

Several researchers have presented approaches to support testing QRs. The QUPER model helps development organizations set appropriate quality targets in a market-driven context, and Berntsson Svensson and Regnell proposed adding also test results to the model to support verification of QRs [7]. Shahrokni and Feldt developed the framework ROAST [32] to support elicitation of testable QRs. ROAST supports refinement of QRs from high level goals to a set of testable requirements, but the framework focuses on robustness requirements and is not applicable to QRs in general. Along the same lines, the same Shahrokni and Feldt also presented RobusTest, another technical framework supporting automated testing of a system's robustness properties by generating JUnit test cases [33]. Cleland-Huang *et al.* introduced a concept of "virtual plumblines" to monitor how a system conforms to quality goals throughout software evolution [12]. By specifying the plumblines as quality assessment models, carefully distributed in the software system, they could then be reevaluated when a system is changed to verify that quality has not deteriorated.

## 3 Case description

We study a large software project at a governmental agency in Sweden. The first author has extensive experience as a consultant at the agency, as also reported in our publication [28]. In this paper we return to the same case, but particularly focus on QRs. In line with the previous publication, we refer to the governmental agency as GOV. GOV administrates subsidies from the European Union within a specific area. The software development organization within GOV develops a new customized platform, offering end-users improved subsidies management using a combination of cloud solutions.

The new platform will combine several existing systems, integrating all steps of the application process into a system-of-systems [3], from the individual request to the final outgoing payment. The project has the policy to use open source software when feasible, utilizing solutions such as Java, JBoss, and PostgreSQL. The runtime environment for several parts of the SoS is Red Hat Linux.

Figure 1 shows an overview of the integration platform under development. Instead of having all 12 partaking systems communicate directly with each other (a) in Figure 1, an integration platform approach is used (b), which is detailed in (c). Each of the 12 systems, as well as each integration between a system and the integration platform, are specified in a separate requirements document.

Two of the most important quality attributes for the new SoS are *interoperability* and *performance*. The SoS comprises multiple dependencies, and most external interfaces will be connected through a common integration platform (cf. b) in Figure 1). Reliable integration with services that provide data is essential, thus providing high quality APIs is an explicit requirement on the SoS. Also, as the number of future users is high, approximately 500 simultaneous users during the peak periods, the network performance is critical. Additionally, the data validations can sometimes require extensive resources, which can affect the performance of the IS, e.g., response times.

We define two key stakeholders of the SoS. *End users* are domain experts working at either GOV or county administrative boards across Sweden. They have detailed knowledge about the application process and assist clients. *Clients* (i.e., individuals and enterprises active in the sector) use the SoS to apply for subsidies. Furthermore, we define the *business* as the organization at GOV that will use the SoS to fulfil their operational needs. The IT department of GOV employs *business analysts*, i.e., requirements engineers responsible for elicitation and specification of the requirements, as well as *software developers* and *system architects*.

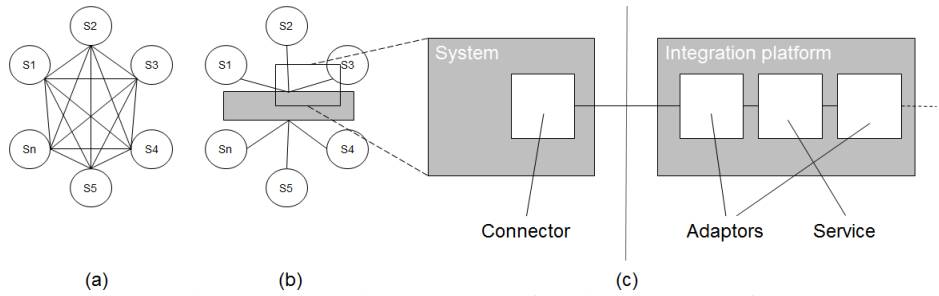

Figure 1. A principle overview of the integration platform.

Analogous to other governmental agencies in Sweden, new development as well as operations at GOV is governed by framework agreements with subcontractors. Several subcontractors develop a large fraction of the SoS together, and also perform RE and testing activities. The development organization employs approximately 100 development engineers in total. Development of the new SoS is organized as 12 different projects, reflecting the number of systems to be integrated, employing roughly 10-20 engineers per project.

The development process at GOV is based on the Rational Unified Process complemented by agile practices expressed as six goals: 1) End users and the business work closely together, 2) High quality through continuous integration and automated testing, 3) Incremental development enables frequent acceptance testing, 4) Continuous delivery of system documentation, 5) Development teams are cross-functional with integrated competences, and 6) Retrospective meetings at every sprint planning. GOV uses the FURPS+ model to elicit and specify QR [25].

Planning within the development project at GOV is performed on three different levels. Between 6-12 months, strategic planning, which involves resource allocation and specification of dates for integration, referred to as milestones. Operational planning (planning which covers more than 3 sprints). Finally, sprint planning (2 weeks) deals with detailed planning of what is to be delivered in the sprint. Further details on the development practices at GOV are reported in our previous paper [28].

## 4     Research methodology

This paper primarily documents experiences from the first author. However, we complement the experiences with an analysis of documents at GOV.

### 4.1 Experiences of the First Author

All experiences reported in this paper belong to the first author, and do not necessarily reflect the views of neither Capgemini nor GOV. Larsson is a software engineering consultant specializing in testing, especially test processes and test management. He has worked in software development projects in the public sector for more than a decade in Sweden and Denmark, with extensive experience of public sector development of large information systems. Furthermore, he has consulted in RE, mainly requirements elicitation and analysis. Note however that the experiences shared in this paper mainly reflect the perspective of a test manager.

### 4.2 Document analysis

The archival analysis of this work is dominated by qualitative analysis. The second and third authors independently analyzed project documentation at GOV to find support for the experiences reported by the first author. The analysis of GOV's archival data, also referred to as content analysis, was conducted on a subset of the available project documentation. We studied all available documents describing the development process at GOV, and selected one of the twelve systems in the SoS for in-depth analysis of all parts and requirements. This is complemented by a light-weight analysis of three additional systems where the requirements are sampled. The emphasis on *what* is in the project documents rather than *how* they were created is inspired with the concept of *artifact-based RE* as described by Méndez Fernández and Penzenstadler [30].

### 4.3 Threats to validity

The primary purpose of this paper is to present the first author's personal experiences. Consequently, our paper is subject to the general limitations of experience reports; Generalization from our findings is uncertain. While we do not discuss extrapolating beyond the case under study, it is relevant to question how general the experiences of the first author are among his co-workers. Other engineers at GOV might highlight other challenges, and also prioritize them differently. However, we increase the validity of our findings by independent data source triangulation, i.e., the second and third authors performed document analysis to identify supporting evidence in GOV's project documentation.

## 5 Results and Discussion

This section first presents findings from the archival analysis, and then reports the first author's experiences from testing QRs at GOV.

## 5.1 QR Information Structure and Information Flow

The development project under study encompasses an extensive document structure. Documents at GOV are hierarchically organized in overall development process guidelines and project documentation for the individual systems. The general quality of the documents is high, and GOV adheres to good practices such as careful change management and consistent templates. All documentation is in Swedish.

Figure 2 shows an overview of the information flow related to QRs at GOV, from RE to the left, to testing at the right side. First, the business analysts create the *Quality Requirement Baseline* (QRB), an overview of the required system qualities, and the *use cases* based on 1) the European Union and Swedish *legislation*, 2) the overall *business model* at GOV, and 3) various *process models* describing high-level usage of the system under development. The system architects create the system's *architecture specification*, complemented by a set of *Interface Descriptions* (IDs) for the relevant inter-system collaboration. The output from the business analysts and system architects is used as input for the software developers (cf. the lower part of the figure, not in focus of this study). Apart from the developed software system (cf. the white box), the test manager uses the QRB and IDs as direct input for testing QRs. The use cases are also relevant; the use cases must be understood to setup a realistic test environment, i.e., most QRs cannot be tested in isolation, rather the test manager needs to establish representative test scenarios in which to verify the QRs.

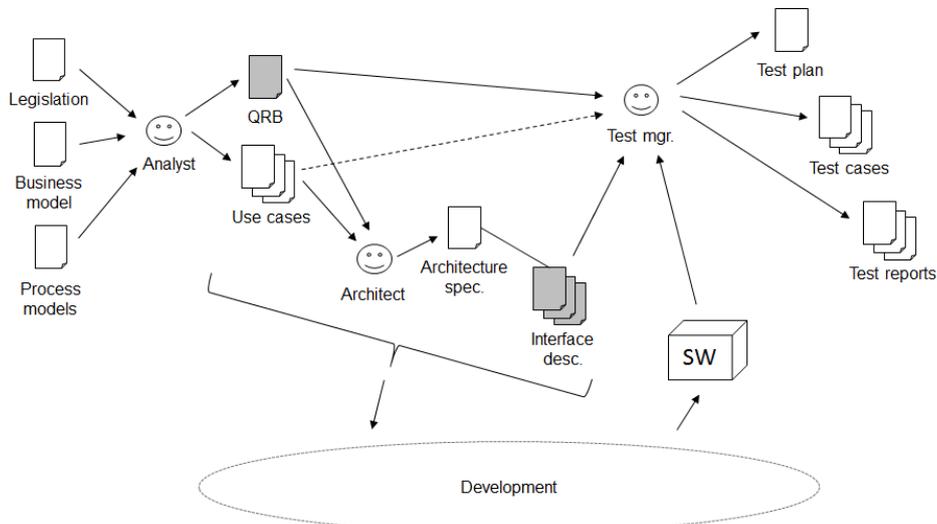

Figure 2. The information flow between RE and testing. Gray documents contain QRs. Solid arrows show direct usage, the dashed arrow shows indirect usage. Based on Stapel and Schneider's work on information flow [35].

As illustrated in Figure 2, two artifact types constitute the key information flow of QRs from RE to testing. First, the *Quality Requirements Baseline* (QRB) is a single document that encompasses the overall qualities of the integrated SoS, i.e., the QRB is the quality centerpiece from an RE perspective. It is structured according to FURPS+, with sections for each corresponding category (see Section 2.1). Examples of system qualities specified in the QRB include: 1) acceptable downtime, 2) recoverability and startup performance, 3) response times, and 4) the number of simultaneous users and actions. Second, the *Interface Descriptions* (ID) refer to a set of documents specifying how the individual systems shall interact within the SoS. Each interface description specifies the behavior of both the producing and consuming end, as well as the transmitted messages. Examples of QRs for the producers include: 1) size and complexity of the messages, and 2) security. Examples of QRs for the consumers include: computational performance, e.g., 1) number of messages processed per second, and 2) time behavior, e.g., time to validate received data.

As a majority of GOV's QRs are specified in the QRB and the IDs, we restrict the discussion in the paper to these two artifact types. This focus mirrors the approach taken by test managers when planning verification of GOV's quality goals. There are about the same number of FRs as there are QRs in the QRB and IDs (note that the 'F' in FURPS+ represents "Functionality"). The FRs are typically high-level textual requirements in natural language, e.g. *"The system should handle payments of environmental support"*. The QRs are written in a similar manner, e.g., *"Response time should not exceed 1 second when searching and registering information on a particular issue"*. While all FURPS+ categories are represented in the QRB usability and implementation/architectural requirements are the biggest (i.e., 'U' and '+' in FURPS+). The level of QR quantification varies depending on type, as reported also in previous work [6]. The reliability and efficiency requirements (i.e., 'R' and 'P' in FURPS+) are often quantified with an explicit target measure, but the other categories are expressed in less precise terms. For the IDs, there is a focus on efficiency and reliability (i.e., 'E' and R in FURPS+). Interestingly, despite specifying interaction within a SoS of information systems, the amount of security requirements is low (security is part of the 'F' in FURPS+).

### 5.2 Experienced challenges in testing quality requirements

The QRB, the central quality document, provides an overall structure to work with QRs at GOV. The categorization of QRs into FURPS+ helps the test managers to identify the key QRs of each category (e.g., usability and reliability) and to plan the verification activities accordingly. Still, testing QRs is far from trivial at GOV.

The IDs, specifying integration within the SoS, are the fundamental artifacts to verify QRs related to the data flows. Early specification of data flows enables early planning of QR verification, but requires cooperation between the system architects and test managers. While some of the IDs reached a stable state early on in the project, others are stabilized later in the process. Hence, as the requirements change, the test plan must be updated to reflect the changes.

Typically, the IDs contain a subset of QR from the QRB, but elaborated for the specific interface. Not all QR from the QRB are applicable though, e.g., usability (as the IDs do not describe any user interfaces). On the other hand, other QRs become more important, e.g., efficiency (as the IDs are specifying integration into a SoS, covering network throughput, response times etc.).

| |
|---|
| Challenge 1: The RE documents evolve while testing is planned and ongoing. |
| Challenge 2: Test managers need to understand the business. |
| Challenge 3: QRs are not quantified. |
| Challenge 4: QRs are not prioritized. |
| Challenge 5: Hard to simulate all operational states. |

Table 2. Main challenges (Ch 1-5) in testing quality requirements at GOV.

Table 2 summarizes our experienced challenges of testing QRs. First, the *QRB and the ID are repeatedly updated during development, thus the test managers do not know when the artifact is stable*. As a consequence, early test plans might target the wrong quality goals, or miss out on verifying entire quality aspects. This confirms previous research, stating that QRs are often discovered late during development [13, 23]. The challenge involved in awaiting a sufficient set of QRs (i.e., enough requirements to enable test planning) also resonates with previous research on dependencies among QRs [4]. Furthermore, the documents are not always properly maintained, i.e., the artifacts do not necessarily reflect the latest quality targets set by for instance the business analysts or the system architect. Inadequate maintenance of requirements documents was also reported as a RET challenge by Bjarnason *et al.* [8].

Second, to plan the verification, *the test managers must have a broad understanding of the business*. The QRs can typically not be interpreted in isolation. Rather, the test manager must understand the domain and possess insights in the data flow through the SoS. This is further complicated by the fact that the QRs are not specified in their context. Rather, they are written as part of the QRB or the ID and in a structure related to the type of requirement rather than their relation to the other requirements and parts of the system. Consequently, the testers must identify and understand the context themselves. In addition, the importance of domain knowledge to software testing is well-known in the literature [20]. At GOV, this challenge is further amplified by the high proportion of consultants involved in software testing; as individual consultants are replaced, important practical know-how might get lost, in line with previous research on risks of tacit knowledge [31].

Third, *QRs are often not quantified*, thus it is difficult to generate appropriate test data. Vague QRs is a frequently reported issue in RE research [6, 19], and without quantification QRs are often not verifiable. Looking at the QRs in the QRB and IDs, performance requirements and sometimes reliability requirements are quantified. Typically with an exact value, rather than a scale. While some types of QRs do not need explicit numbers, e.g., usability and maintainability, having explicit quantification of QRs help to eliminate subjectivity during testing, in line with observations by Chung *et al.* [15]. Indeed, generating test data to verify a software system is a difficult research topic in itself [2], and in conjunction with vague requirements the challenge is further intensified. On the other hand, some of the levels that actually are specified for QRs at GOV are merely early guess estimates that must be updated later, forcing re-planning of testing.

Fourth, *QRs are not prioritized*. As a set of QRs typically involves interdependencies [4, 9, 13], understanding quality trade-offs is fundamental. While there are policies regarding risk assessments at GOV, they are not always followed. Without input on priorities and risks, test plans become more sensitive to changes as mitigation plans are more difficult to create.

Fifth, it is *hard to generate test data that reflects the various operational states* of the future SoS, as required for e.g. load and performance testing. The QRs expressed in the IDs put constraints that require the test organization to simulate various states using stubs and mock objects; different QRs need different operational states to be verified, and verification of certain QRs might require considerable provocation of the SoS. Some states are difficult to create in a lab environment, e.g. user interaction that returns a specific expected result within a given timeframe, simultaneously as real time data is exchanged over some interfaces with other systems. Another example is verification of complex business logic that is run concurrently in parallel activities. The validation of transmitted messages within the SoS requires heavy read and write to databases, which affects response times of systems retrieving data from the same databases. Currently, the QRs do not specify which use cases can run in parallel or what the quality level should be under such constrained conditions, making both design of test cases and evaluation of test results challenging.

### 5.3 Discussion and future research

GOV is already addressing the five reported challenges to some extent. One of the main initiatives is to get different experts and roles to work closer together. Inspired by *integrated RE* [36], there is an ongoing discussion on a more iterative approach to RE. Specifying quantitative targets to QRs before any related source code has been developed often reveals the inferiority of such an approach, as the QRs typically need subsequent updates (Ch 1). The literature emphasizes continuous maintenance of requirements rather than big changes [29], and a constant dialogue between requirements engineers and testers [8], which for GOV could *turn the QRB and IDs into "living documents"*, i.e., artifacts that continuously evolve.

Transferring knowledge of various operational scenarios at GOV to the development project (including the testing activities (Ch 2)) should be facilitated further. While Bjarnason *et al.* state the importance of communication between requirements engineers and testers [8], we highlight that *knowledge sharing between system architects and test managers* is equally important at GOV, as the architecture is fundamental input to testing QRs (cf. Figure 2). The interleaving of RE and architecture, i.e., the twin peaks model [11], has attracted much research, but we argue that *also the test perspective should be represented in the twin peaks model* (resulting in a "triplet peak" model) to support the test managers' system understanding and help specifying testable QRs. A straightforward step toward alignment of RE, architecture, and testing perspectives at GOV could be to schedule additional joint meetings and workshops, especially in the beginning of projects.

One solution proposal from the literature is the QUPER model, an approach to support setting quality targets [7]. In the QUPER model, quality is modelled as a non-linear continuum with specific breakpoints for utility, differentiation, and saturation. By visualizing the quality targets and the different breakpoints, *QUPER could stimulate discussion and quantification of QRs* at GOV (Ch 3). While QUPER was developed for RE in a market-driven context, the model can still support discussions on quality targets for products without explicit competitors. Figure 3 illustrates how QUPER can support setting a target for a response time QR.

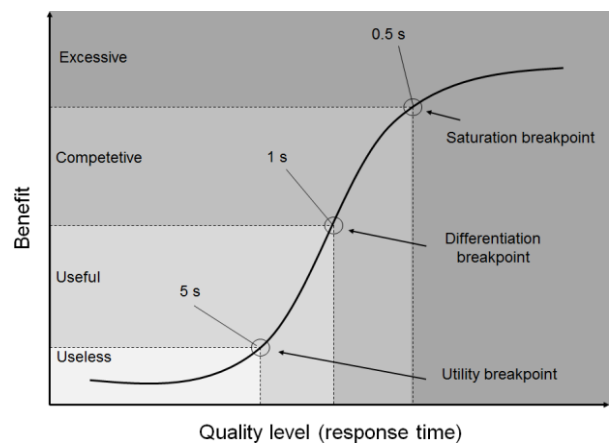

Figure 3. Finding a quality target with the QUPER model. Response times > 5 s represents a useless product, whereas improvements beyond 0.5 s is not recognized by users. A quality target of 1 s is estimated to impress users.

A common trade-off in software engineering is having complete information available when you start an activity versus being responsive to a changing environment. GOV faces the same challenge, as explained in Ch 1: "The RE documents evolve while testing is planned and ongoing" and Ch 4 "QRs are not prioritized". However, not all QRs have an equal impact on the system [10, 16]. Some might be more fundamental and impact several parts, more significantly changing implementation and hence also testing. Typically, QRs have a more profound impact than FRs, but also among the QRs, their impact is different. Chen *et al.* introduced the concept of *architecturally significant requirements* to describe this phenomenon [10]. However, the overall problem to identify the significant requirements is not well researched and it is specifically not clear how to identify the significant QRs [10]. We would like to further investigate techniques and methodologies to identify "test significant" QRs. The activity should focus and prioritize the requirements that are the most critical to fully specify early in the project life-cycle.

Verification of some QRs at GOV requires monitoring various aspects of the SoS over time. However, interpreting the results of monitoring is still not a simple task. Performance and load testing of an interface not only require monitoring of both the producing and consuming sider; As the business flows gets more complex, the number of interfaces grows, thus monitoring of other applications must also be set up

(related to Ch 5). For example, monitoring of the database server has proven to be fundamental. Especially in the iterative integration of more systems into the integration platform, values need to be monitored and regression tests run not once but several times as the system changes. Another possibility to verify QRs over time at GOV could be to *implement "virtual plumblines"* as suggested by Cleland-Huang *et al.* [12]. In the GOV context, this could for example be used to continuously monitor performance goals as the system is developed; when a certain system quality deteriorates, such an alarm system could provide the test leader with early warnings. By employing practices from integrated RE [36] together with data monitoring tools, GOV hopes to also see improvements in the area of Ch 1 (changing requirements) and Ch 2 (business understanding of test managers).

## 6   Conclusion

This paper reports the first author's experiences from testing QRs at a large governmental SoS development project in Sweden. We highlight five challenges involved in testing QRs, and complement the experiences by supporting evidence from an analysis of project documentation. The five main challenges in testing QRs at GOV are: 1) RE documentation evolves during testing, 2) test managers need extensive understanding of the business, 3) QRs are not quantified, 4) QRs are not prioritized, and 5) simulating all operational states is difficult, challenging testing prior to deployment.

To help GOV tackle these challenges, we match them with solution proposals from the research literature. Evolving QRs is a well-known challenge in software engineering, and could be addressed by an integrated RE approach, i.e., plan for RE to continue throughout the entire development project (Ch 1). Strengthened communication between the stakeholders from the "twin peaks" (RE and architecture) and the test managers would support RET knowledge transfer and specification of testable QRs (Ch 2). Moreover, improved communication among roles, e.g. business, architects and test managers, could help identifying architecturally significant requirements, e.g., to prioritize QRs that have considerable impact on the test planning (Ch 4). The QUPER model could be a tool to stimulate discussion on quality levels and support quantification (Ch 3). GOV already monitors several quality measures, but the process could be further inspired by "virtual plumblines" to quickly detect quality deterioration (Ch 5).

Future work could systematically explore which of the solution proposals GOV considers most promising to support testing QRs, and also try to implement and evaluate the corresponding process improvements.